# Parallel Proportional Fusion of Spiking Quantum Neural Network for Optimizing Image Classification

Zuyu Xu[1], Kang Shen[1], Pengnian Cai[1], Tao Yang[1], Yuanming Hu[1], Shixian Chen[2], Yunlai Zhu[1], Zuheng Wu[1], Yuehua Dai1, Jun Wang[1*], Fei Yang[1*]

[1]School of Integrated Circuits, Anhui University, Hefei, Anhui, 230601, China.
[2]School of Computer and information, Anhui Normal University, Wuhu, Anhui, 241003, China.

E-mail(s): iamwj7@163.com; feiyang-0551@163.com;

**Abstract**

The recent emergence of the hybrid quantum-classical neural network (HQCNN) architecture has garnered considerable attention due to the potential advantages associated with integrating quantum principles to enhance various facets of machine learning algorithms and computations. However, the current investigated serial structure of HQCNN, wherein information sequentially passes from one network to another, often imposes limitations on the trainability and expressivity of the network. In this study, we introduce a novel architecture termed Parallel Proportional Fusion of Quantum and Spiking Neural Networks (PPF-QSNN). The dataset information is simultaneously fed into both the spiking neural network and the variational quantum circuits, with the outputs amalgamated in proportion to their individual contributions. We systematically assess the impact of diverse PPF-QSNN parameters on network performance for image classification, aiming to identify the optimal configuration. Numerical results on the MNIST dataset unequivocally illustrate that our proposed PPF-QSNN outperforms both the existing spiking neural network and the serial quantum neural network across metrics such as accuracy, loss, and robustness. This study introduces a novel and effective amalgamation approach for HQCNN, thereby laying the groundwork for the advancement and application of quantum advantage in artificial intelligent computations.

1. Introduction

Image classification is a fundamental domain in the field of deep learning [1]. Despite its widespread application, the expanding size of training datasets and the growing complexity of models have pushed deep learning algorithms, heavily reliant on substantial computational power, to the limits of classical computers' processing capacity [2]. The exponential growth in data and model complexity has underscored the urgency for innovative solutions that surpass current computational constraints. In addressing the challenges posed by this surge in demand, there is a crucial need for developing highly efficient models capable of adeptly processing extensive datasets to achieve precise and swift image classification. The current landscape underscores the critical necessity for breakthroughs in model architecture and computational strategies to ensure that image classification remains both accurate and computationally feasible amid growing data and complex model structures. In response to these imperatives, researchers are compelled to explore novel avenues promising enhanced efficiency, pushing the boundaries of what is achievable within the realm of classical computing for image classification tasks.

Recently, quantum computing algorithms, owing to their parallelism in handling information, have emerged as a promising computing paradigm that could address the challenges mentioned above [3-6]. The integration of quantum algorithms with deep learning models has the potential to enhance computational capabilities while preserving model accuracy [7]. Presently, numerous deep learning techniques employing quantum algorithms have been developed to enhance computational efficiency. For instance, the quantum perceptron computation model merges quantum theory with neural networks, introducing a singular quantum circuit that simulates the nonlinear input-output function of a classical perceptron with $O$(N) parameters for system size N [8]. By decoupling the feature extraction process from the deep learning model, it aims to reduce the complexity of training. It utilizes quantum circuits to extract features through quantum measurements, generating a quantum representation of the image. This quantum representation is then employed to train a fully connected neural network for image classification [9]. Nevertheless, relying solely on quantum implementation for neural networks poses challenges, particularly when confronted with real-world datasets of considerable size. This mapping swiftly becomes impractical for contemporary quantum computers equipped with fewer than 200 qubits [10].

More recently, a hybrid framework known as Hybrid Quantum-Classical Neural Networks (HQCNN) has been proposed for image classification. In this framework, the quantum component is represented by a parameterized circuit designed to extract essential image features, while the classical counterpart performs the classification. HQCNN has demonstrated effectiveness in various image classification tasks, outperforming its classical counterpart and exhibiting enhanced generalizability and robustness across multiple applications. Notably, quantum adaptations of deep learning techniques, such as convolutional neural networks[11], Boltzmann machines[12], generative adversarial networks[13], and quantum transfer learning[14], have been devised for a wide range of image classification tasks. The Quantum Superimposed Spiking Neural Network (QS-SNN), drawing inspiration from quantum mechanics and brain phenomena, is adept at handling inverted color images of backgrounds [15]. Another hybrid model, the Spiking Quantum Neural Network (SQNN), combines classical and quantum elements in a spiking feedforward neural network. This model is employed to tackle challenging image classification tasks, particularly in the presence of noise and adversarial attacks [16]. However, most of the current hybrid spiking quantum network studies for image classification adopt relatively small-scale, implementable quantum circuits and classical neural networks working in series, passing information from one network to another, which will introduce information bottlenecks in the representation capability of the model, severely limiting the expressivity and trainability of the network [17].

The exploration of parallel structures in classical networks has been a subject of extensive study, exemplified by notable architectures like Google Inception [18], which has substantially enhanced network performance. Motivated by these advancements, in this work, we present a parallel proportional fusion of the spiking quantum neural network (PPF-SQNN) for the image classification. Within this framework, we exploit the capabilities of spiking neural networks and quantum neural networks to autonomously extract image feature information. Subsequently, we merge the extracted information, considering diverse dimensions and orientations. This fusion process enhances the data, imbuing it with increased dimensionality, diversity, and stability. By integrating these components, our approach enables a more robust and comprehensive capture of dynamic features inherent in images. This improved representation ultimately enhances the classification capacity, particularly in addressing the intricacies of complex images. The proposed hybrid quantum-spiking neural network demonstrates the potential to advance image classification methodologies, with implications for various applications, ranging from artificial intelligence to computer vision.

The contributions of this paper are as follows: (1) A hybrid PPF-SQNN with is proposed to obtain more comprehensive and accurate image feature information, and improve the expression and computing ability of massive data. (2) An innovative network weight optimization algorithm is proposed. (3) The network is parameterized and compared with other neural networks, proving that the network has advantages in image classification. (4) Noise test is carried out, and PPF-SQNN is found to have extremely high noise immunity.

The rest of this paper is organized as follows. Section 2 provides the basic idea and theoretical knowledge of PPF-SQNN, in which Section 2.3 introduces a network weight optimization algorithm. Section 3 conducts a series of parameter optimization experiments on

PPF-SQNN and compares it with other neural networks, proving that the network has advantages in image classification. Section 4 conducts noise tests on the network. Section 5 discusses the generalization of the network. Section 6 summarizes the work.

2. The Proposed PPF-QSNN Architecture

In the realm of image classification tasks, we present a pioneering approach through the development of a hybrid quantum spiking neural network (QSNN) architecture, featuring a parallel proportional fusion (PPF) strategy as depicted in Figure 1. Our primary aim is the seamless integration of the distinctive capabilities of a spiking neural network (SNN) and a variational quantum circuit (VQC), facilitating the extraction of comprehensive image feature information from both classical and quantum perspectives. The spiking neural network demonstrates high efficiency and rapid processing of classical information, while the quantum neural network excels in capturing more nuanced feature information. Importantly, we integrate these two parallel networks and fine-tune the quantum and classical proportions of the output segments to explore the optimal impact on image classification. This strategic design of the fusion architecture enables a comprehensive analysis of image data from multiple dimensions, thereby enhancing the classification capability, especially for complex images.

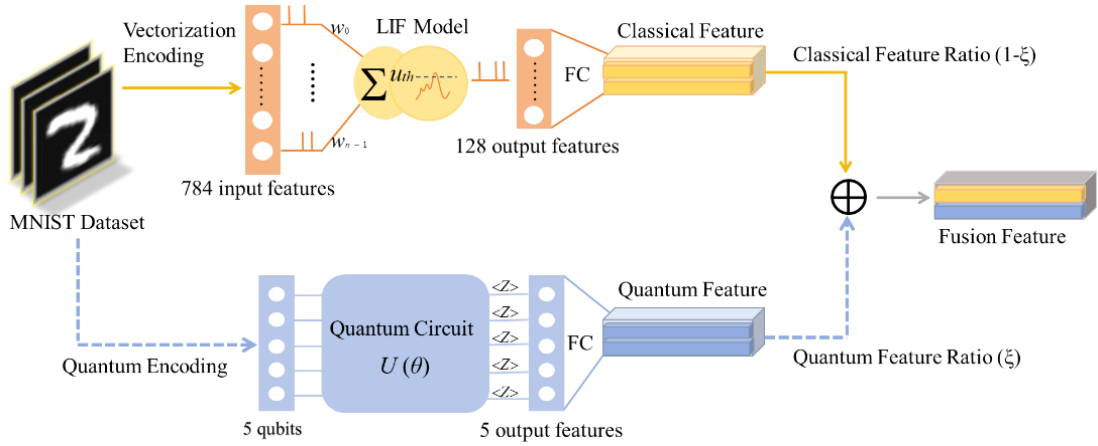

**Figure 1:** The architecture of the hybrid quantum spiking neural network is presented, featuring a parallel proportional fusion strategy. Initially, image pixels undergo preprocessing, and the processed data is then inputted into both the spiking layer and the quantum circuit for comprehensive feature extraction. The subsequent step involves proportional fusion of feature information from both sources, culminating in the formation of the composite data crucial for image classification. The symbol $U(\theta)$ represents a series of quantum unitary gates encoding phase, while $<Z>$ denotes quantum measurement performed on the qubits.

2.1 Feature extraction in spiking neural networks

In our image experiments, we employ a Spiking Feedforward Neural Network (SFNN) [19], which exclusively utilizes spike events as input. During the data preprocessing stage, image pixels undergo initial reshaping into a vector, followed by conversion into a series of spike streams. These spike streams are then input into the SFNN's input layer. The conversion from each image pixel to a spike stream is accomplished using the Poisson distribution method [20]. Notably, the firing rate of the spike stream is directly proportional to the image intensity.

our work leverages the Leaky Integrate-and-Fire (LIF) model [21], a successful and straightforward method for modeling SFNN. The LIF model is mathematically represented by the differential equation:

$$\tau m \frac{dv(t)}{dt} = -(v(t) - Vrest - R_m I_n(t)) \ , \qquad (1)$$

Where $v(t)$ is the membrane potential of a single neuron, $V_{rest}$ is the resting membrane potential and the membrane time constant $\tau_m$ is given by the membrane resistance $R_m$ multiplied by the membrane capacitance C. $I_n(t)$ denotes the total input from presynaptic neurons.

The fundamental architecture of a SFNN comprises two layers consisting of synapses and spiking neurons. Over an observation period, synapses receive a sequence of input stimuli, transforming them into spikes. Spiking neurons are intricately connected to synapses, with each connection possessing an adjustable weight. Throughout this process, the membrane potential of a spiking neuron is the weighted sum of spikes on synapses. The internal potential of the i-th neuron can be denoted as:

$$v_j(t) = \sum_i w_{ji} K_i(t-\tau) + V_{rest}, \qquad (2)$$

Where $w_{ji}$ represents the connection weight between the *i*-th neuron and the *j*-th presynaptic neuron, and $K_i(t-\tau)$ represents the spike train of the *i*-th presynaptic neuron. When the membrane potential of the *j*-th neuron exceeds the firing threshold ($\theta_j > 0$), the neuron emits an output spike using a time-based encoding.

For image classification problems, the ReLU (rectified linear unit) activation function [22] is widely used for the intermediate feature processing of the hidden layer. Compared with sigmoid/tanh functions, ReLUs greatly accelerate the convergence speed of stochastic gradient descent due to their linear and unsaturated form [23]. We employ ReLUs in the hidden layers during our training. A typical mathematical description is given as

$$x_i = \max(0, \sum_j w_{ij} x_j), \qquad (3)$$

Where $x_i$ is the activation of the unit, $w_{ij}$ is the weight connecting the unit *j* in the previous layer to the unit in the current layer, and $x_j$ is the activation of the unit *j* in the previous layer. By continuously updating all the activations of the current layer based on the activations of the previous layer, the input propagates through the network to activate the output label neuron.

In the training phase employing standard error backpropagation, the error gradient undergoes a layer-by-layer backward propagation process. This facilitates the iterative adjustment of individual weights, minimizing errors through the implementation of stochastic gradient descent [22]. Within these networks, the training procedure dynamically tunes the randomly initialized weight matrices describing the intricate connections between layers, ultimately working towards the minimization of the overall error through stochastic gradient descent. This training methodology plays a pivotal role in refining the SNN's capacity to effectively capture and learn complex patterns within the provided data, thereby enhancing its classification and decision-making capabilities.

2.2 Variational Quantum Circuit

A Variational Quantum Circuit (VQC) is a versatile framework within the realm of quantum computing that holds significant promise for various applications, particularly in quantum machine learning and optimization tasks [24-26]. The whole VQC process consists of encoding, variation, and measurement elements.

In the process of encoding, our methodology takes inspiration from the Flexible Representation of Quantum Images (FRQI) method [27]. This innovative approach involves the normalization and dimensional reduction of the entire image before encoding. Given an image with pixels $\theta = (\theta_0, \theta_1, \theta_2, ..., \theta_{2^{2n}-1})$, the pixels of which have been normalized so that $\theta_i \in [0, \frac{\pi}{2}]$, $\forall i$, the encoding state is:

$$|\psi(\theta)\rangle = \frac{1}{2^n} \sum_{i=0}^{2^{2n}-1} (\cos\theta_i |0\rangle + \sin\theta_i |1\rangle) |i\rangle, \qquad (4)$$

Where $|\psi(\theta)\rangle$ represents the quantum encoded image, $\cos\theta_i|0\rangle+\sin\theta_i|1\rangle$ encodes the color of the pixel, while $|i\rangle$ encodes the location of the pixel in the image.

FRQI satisfies the quantum state constraint in Equation (5):

$$\||\psi(\theta)\rangle\| = \frac{1}{2^n}\sqrt{\sum_{i=0}^{2^{2n}-1}(\cos^2(\theta_i)+\sin^2(\theta_i))} = 1, \tag{5}$$

Subsequently, a sequence of variational and fixed quantum gates is implemented in the circuit. Variational gates may include Pauli rotation gates, each necessitating a single-qubit time-evolution Hamiltonian specific to its corresponding Pauli gate. Meanwhile, fixed quantum gates can encompass controlled-NOT gates and Hadamard gates.

The variational and fixed quantum gates employed in our implementation are visually represented in Figure 2. Hadamard gates (H gate) are utilized as the input on the ground state, generating an equal superposition of quantum states and, consequently, positioning the system away from both poles of the Bloch sphere. The Hadamard gate is mathematically described by the matrix [28],

$$H = \frac{1}{\sqrt{2}}\begin{pmatrix} 1 & 1 \\ 1 & -1 \end{pmatrix}, \tag{6}$$

Single-qubit rotation gates $R_Y(\theta)$ are adopted to encode rotations along y-axis [29]:

$$R_Y(\theta) = \exp(-\frac{i\theta}{2}Y) = \begin{bmatrix} \cos\frac{\theta}{2} & -\sin\frac{\theta}{2} \\ \sin\frac{\theta}{2} & \cos\frac{\theta}{2} \end{bmatrix}, \tag{7}$$

We use CNOT gates in the training process to entangle the quantum states of each qubit. Phase information is encoded by the combination gates $R_\omega(\theta)$ in the quantum layer. The training of quantum circuits is based on the output spikes of SNN block, which are given as trainable parameters $\theta$ in quantum circuits.

$$R\omega(\theta) = R_ZR_XR_Z = \exp(-\frac{i\theta}{2}Z)\exp(-\frac{i\theta}{2}X)\exp(-\frac{i\theta}{2}Z) = \begin{bmatrix} \exp(-\frac{i\theta}{2}) & 0 \\ 0 & \exp(\frac{i\theta}{2}) \end{bmatrix}\begin{bmatrix} \cos\frac{\theta}{2} & -i\sin\frac{\theta}{2} \\ -i\sin\frac{\theta}{2} & \cos\frac{\theta}{2} \end{bmatrix}\begin{bmatrix} \exp(-\frac{i\theta}{2}) & 0 \\ 0 & \exp(\frac{i\theta}{2}) \end{bmatrix}, \tag{8}$$

The last step is to perform Pauli-Z($\sigma_z$) measurement[9] on the qubit to obtain the quantum measurement-based features. Let the state of the single qubit before measurement be $|\varphi\rangle$, where

$$|\varphi\rangle = \frac{1}{\sqrt{\alpha^2+\beta^2}}\begin{pmatrix} \alpha \\ \beta \end{pmatrix}, \tag{9}$$

After Pauli-Z($\sigma_z$) measurement, the quantum bit state $|\varphi\rangle$ collapses, and the obtained value is:

$$Tr(|\varphi\rangle\langle\varphi|\sigma z) = \frac{|\alpha|^2-|\beta|^2}{|\alpha|^2+|\beta|^2}, \tag{10}$$

Where $Tr$ denotes the trace operation.

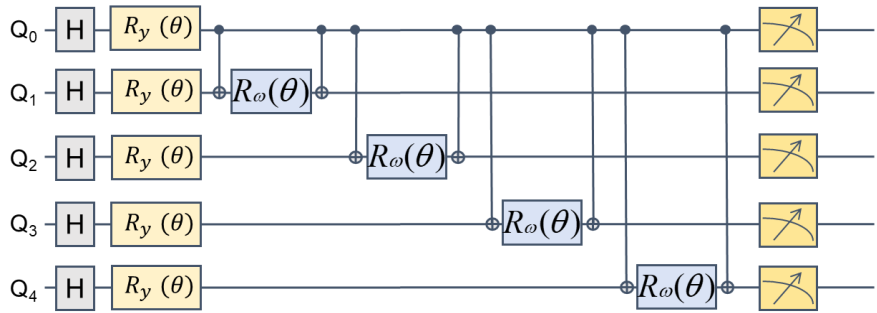

**Figure 2:** Quantum circuit diagram. There are 5 quantum bits $Q_0$~$Q_4$ in total, and $R_\omega(\theta)=R_ZR_XR_Z$ combination gates are used to encode the phase information.

2.3 Optimization Algorithm for Network Proportion

In our architecture, feature outputs derived from the spiking neural network and the quantum circuit undergo uniform processing individually to ensure dimensional compatibility. This enables their seamless integration to generate a novel fusion feature vector, encapsulating the event-driven dynamics of the spiking neural network and the quantum superposition and entanglement characteristics intrinsic to the quantum neural network. The ensuing fusion feature vector serves to comprehensively and diversely represent image information. Subsequent to this, we provide a detailed exposition on the feature output processing.

For image classification tasks, the widely employed softmax activation function, prevalent in deep artificial neural networks, is utilized for processing the output from the spiking layer. The decision on whether a neuron should spike is determined by calculating the softmax of the membrane potential. The ultimate classification outcome, during the stimulus presentation, corresponds to the index of the neuron exhibiting the highest firing rate, as elaborated below (Equation 11), where denotes the feature vector, and and represent its constituent elements.

$$\mathrm{soft\,max}(Q_i) = \frac{\exp(Q_i)}{\sum_j \exp(Q_j)}, \qquad (11)$$

Where $Q$ is the feature vector, and $Q_i$ and $Q_j$ are elements in it.

In the VQC, the measurement of the quantum state leads to its collapse into a classical state. The subsequent linear combination of the measurement result and neuron weights is facilitated through a linear layer. This mechanism enables the neural network to dynamically adjust the activation levels of neurons based on the measurement outcome. The feature outputs are further processed using the softmax function.

The softmax function functions to map feature outputs to a probability distribution, converting the output of each feature into a probability value within the range of 0 to 1. The sum of probabilities across all feature outputs totals 1.

Denoting the classical feature vector obtained after softmax processing as $Q_c$, the quantum feature vector as $Q_q$, and the resulting fused feature vector as follows $Q_h$:

$$Q_h = \xi Q_q + (1-\xi)Q_c, \xi \in [0,1], \qquad (12)$$

where $\xi$ is the quantum proportion coefficient, the number of elements in the vector $Q_h$ is the same as the number of image classification categories, and the element with the largest value is the prediction result.

To quantitatively assess the model's performance and prediction quality, we employ the negative log-likelihood loss function for calculating the model's loss value. Rooted in maximum likelihood estimation, this function transforms probabilities into losses through a negative logarithmic operation. The model's predicted probability distribution is then compared with the one-hot encoding of the actual label, as expressed mathematically:

$$Loss = -\frac{1}{N}\sum_N \log p(y_i | x_i), \qquad (13)$$

Where $N$ represents the number of samples, each sample denoted by $x_i$ with the corresponding real label $y_i$.

During training, minimizing the loss value is achieved by utilizing the negative log-likelihood loss function. The backpropagation algorithm is employed to update the model parameters, enhancing its ability to approximate the distribution of real labels.

3. Experiment
3.1 Dataset

In order to verify the feasibility of the model, we use the handwritten dataset "MNIST" for the image classification task. MNIST[30] is a benchmark dataset of handwritten digits, which consists of 60000 training images and 10000 test images, composed of handwritten images of 10 classes from 0 to 9. Each sample is a 28 × 28 pixel grid and has been widely used in the SNN literature[31-33], as shown in Figure 3. In order to test the model's processing effect on a large dataset, we chose to perform a 10-classification task directly.

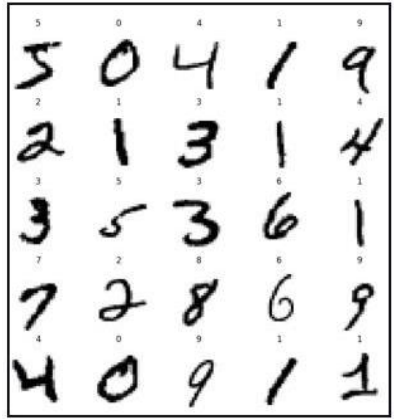

**Figure 3:** MNIST Handwritten Dataset Example Random Samples Images.

3.2 Parameters Optimization

In this investigation, we rigorously explore pivotal factors that influence the experimental outcomes of image classification tasks, namely the quantum ratio coefficient and the number of qubits. Our aim is to offer a comprehensive understanding of how these parameters influence the overall performance of the network.

The intricate process initiates with expanding all features within each sample into a 784-length vector. This vector serves as the input to the spiking neural network, undergoing a nonlinear transformation via the Spiking ReLU function. Subsequently, the TemporalAvgPool layer facilitates data averaging. The pooled data is then directed to a linear layer, generating a classical feature vector $Q_c$. Simultaneously, the data undergoes dimensionality reduction before being transmitted to the quantum circuit for feature extraction and quantum measurement, resulting in a quantum feature vector $Q_q$. The fusion of these two feature vectors is accomplished through a stitching process outlined in formula (12), yielding a feature vector characterized by diversity, stability, and high dimensionality. The number of elements in this feature vector corresponds to the number of image classifications, with each element representing the probability of a specific category. The category with the highest probability serves as the network's classification for the given image.

Equation (12) reveals the dynamic nature of the fusion feature vector's quantum and classical information proportions, contingent upon different values of ξ. To comprehensively assess the influence of varying quantum ratios on classification accuracy, we systematically compute accuracy across a range of quantum ratios, including ξ = 0 to 1, as depicted in Figures 4(a) and 4(b). To mitigate the risk of overfitting, accuracy calculations are performed during both the training and testing phases. The outcomes delineate notable fluctuations in accuracy corresponding to different quantum proportions, with the optimal quantum proportional coefficient identified as 0.8, as illustrated in Fig. 4(c). At this quantum ratio, the training accuracy for image classification reaches 97.1% with a negligible error bar of 0.15%. These findings indicate that the fusion of quantum and classical networks in parallel, according to a specific proportion, enhances the effectiveness of image classification.

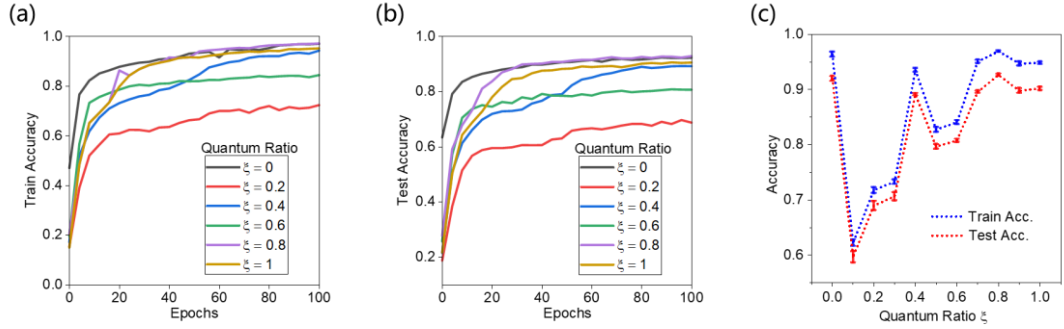

Figure 4. Comparison of accuracy across various quantum proportion coefficients. (a) Training accuracy curves. (b) Testing accuracy curves. (c) Training and testing accuracy versus the quantum proportion coefficients, the error bars are calculated through ten measurements.

In addition to the quantum proportion coefficient, the number of quantum bits emerges as another crucial parameter. Quantum computing, grounded in the principles of quantum mechanics, relies on quantum bits as its fundamental units. The addition of each quantum bit results in an exponential increase in the computational system's complexity. Therefore, determining the optimal number of quantum bits is imperative for tackling more intricate computational tasks. Initial tests assessed the accuracy under various quantum proportion coefficients for qubits = 5, 6, and 7, as depicted in Figures 5(a) and 5(b). The results established that the optimal quantum proportion coefficient of 0.8 remains unchanged with fluctuations in the number of quantum bits. This consistency underscores the robustness of the quantum proportion coefficients for our proposed PPF-SQNN.

Subsequently, we conducted a comprehensive analysis of the accuracy across various iterations for each quantum bit, as depicted in Figure 5(c) and (d), while concurrently tracking the evolution of negative log-likelihood (NLL) loss values throughout the training and testing phases. The outcomes of these experiments underscore the model's remarkable insensitivity to the number of quantum bits, indicative of its capacity to sustain consistent performance within the intricate landscape of quantum computing. This robustness to the quantum bit count offers dual-fold advantages. Firstly, it attests that our model does not excessively lean on a multitude of quantum bits for achieving optimal performance, thereby signaling heightened robustness and reliability. Secondly, given the current embryonic state of quantum computing, conventional computing platforms maintain a relative maturity. The flexibility of our model, unconstrained by the quantum bit count, positions it for widespread applicability across existing traditional computing platforms, whether in the training or inference stages. Furthermore, a notable trend emerges during the training and testing stages, wherein the NLL undergoes a gradual decrease. This trend is not only indicative of the model's learning efficacy but also signifies the realization of the expected optimization effect. As we navigate the dynamic landscape of quantum computing, these findings not only contribute valuable insights into the adaptability and resilience of our model but also advocate for its practical viability across the existing spectrum of computational platforms.

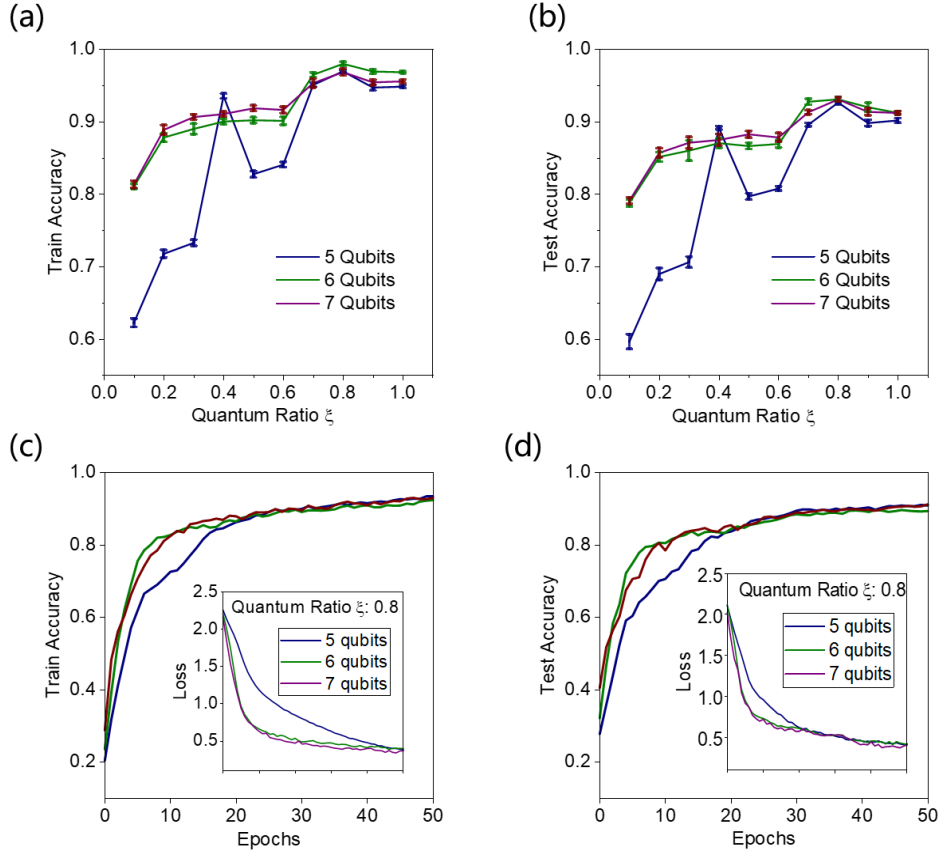

Figure 5. Accuracy under different qubits. (a), (b) Training and testing accuracy versus the quantum proportion coefficients for different qubits. (c), (d) Training and testing curves with the quantum ratio of 0.8 for different qubits.

3.3 Network Performance Comparison

To comprehensively assess the effectiveness of our proposed network architecture, we selected two benchmark models, the classical SFNN [19] and SQNN [16]. The SQNN serves as a representative example of serial hybrid classical quantum neural networks, renowned for its comparative advantages in diverse applications. Our evaluation aimed to provide an in-depth comparison by scrutinizing the accuracy and confusion matrix of these three networks. The networks' accuracy was scrutinized across various iterations to capture their performance dynamics. The elucidating results, presented in Figure 6, accentuate the exceptional capabilities of our network.

Figure 6(a) indicates that the SFNN exhibits the fastest learning rate, while the SQNN displays a slower learning rate. It is noteworthy that faster training can enhance computational efficiency but may lead the model to overshoot the optimal weight. Conversely, slower training rates can contribute to improved generalization, providing the model with more opportunities to discern subtle patterns in the data [34]. However, this deliberative approach may also result in the model getting trapped in local minima, hindering exploration of other regions within the parameter space. Our proposed PPF-SQNN demonstrates a moderate training rate and it also exhibits remarkable proficiency, achieving a commendable training accuracy of 97.1% in the classification task, outperforming both SFNN 96.5% and SQNN 94.1%. Here, the quantum proportional coefficient is set at 0.8. Importantly, during the testing phase, our network sustains its excellent performance, surpassing the testing accuracy of both SFNN and SQNN. This experiment underscores the superior capability of our network in handling large-scale datasets, thereby offering substantial advantages for classification problems.

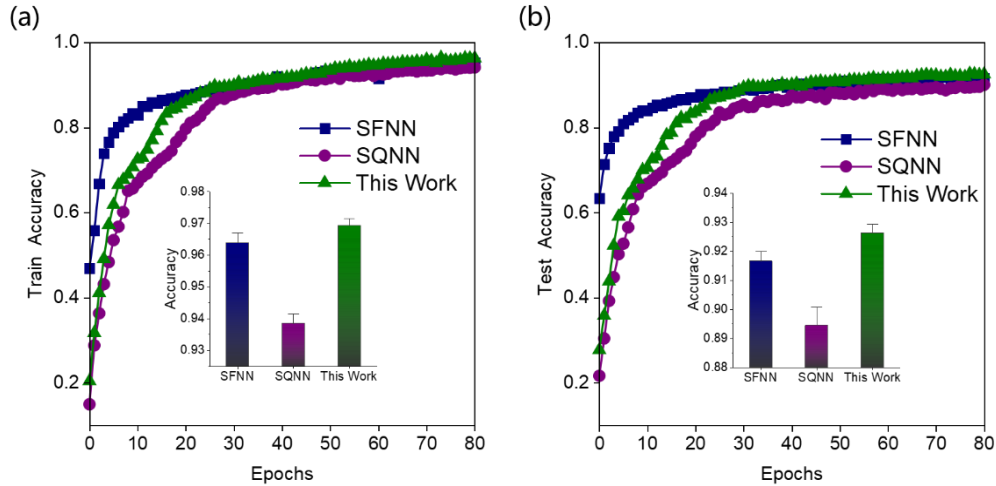

**Figure 6:** Comparison of accuracy between SFNN, SQNN, and our proposed network, the error bars are calculated through ten measurements.

We conducted an extensive performance evaluation of each model by scrutinizing their confusion matrices, as illustrated in Figure 7. The confusion matrix, a widely embraced evaluation tool for multi-class classification problems, surpasses simple accuracy metrics by providing a nuanced analysis of classification accuracy, thereby unraveling the intricate performance characteristics of the models. In this comparative analysis, our network emerged as notably more reliable and stable, reinforcing its discernible advantages. The confusion matrix, serving as a reflection of a model's prediction accuracy across diverse categories, unveiled key insights into the comparative performance of each model. Notably, our network exhibited superior performance in critical categories, attesting to its enhanced classification capabilities and its ability to discern feature differences more accurately across varied categories.

A deeper dive into the specifics of the confusion matrix shed light on our network's adeptness in handling ambiguous categories. Demonstrating a remarkable ability to discern subtle variations between similar categories, our network adeptly allocated samples to their respective categories. This heightened capability is ascribed to the intricate deep structure and sophisticated feature extraction mechanism embedded within our network, enabling it to learn and leverage crucial features within the data for more precise classification. It is paramount to underscore the lower misclassification rates depicted in our network's confusion matrix. This signifies a reduction in classification errors across diverse categories, underscoring the network's heightened reliability and robustness in distinguishing input samples from different categories. Such results underscore the stability and high reliability of our network in effectively addressing classification problems.

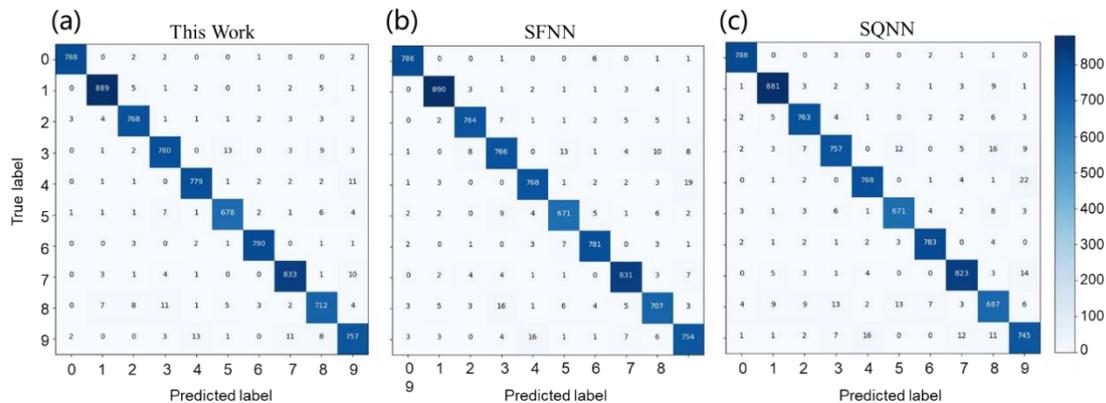

**Figure 7:** Comparison of Confusion Matrices.

## 4. Noise Testing
### 4.1 Uniform Noise and Gaussian Noise

To thoroughly evaluate the robustness of our network, we chose two prevalent types of noise: Uniform noise and Gaussian noise. These noise variations play a pivotal role in our experiments, facilitating a nuanced assessment of the network's performance across a spectrum of noise levels and types, thereby shedding light on its adaptability to diverse scenarios.

Uniform noise adheres to a uniform distribution, evenly spreading noise values within a predefined range. Its introduction injects randomness into the data, thereby augmenting the complexity of sample features. The consequential impact of incorporating uniform noise into the data is visually depicted in Figure 8(a). Conversely, Gaussian noise follows a normal distribution, with noise values centered around a mean and exhibiting a specific standard deviation. The introduction of Gaussian noise authentically simulates real-world randomness and variability, as depicted in Figure 8(b).

Through a meticulously designed set of experiments involving both uniform and Gaussian noise, we conducted a thorough evaluation of our network's robustness. This comprehensive analysis provides valuable insights into the network's adaptability, offering a foundation for enhancing its performance in complex and dynamic environments.

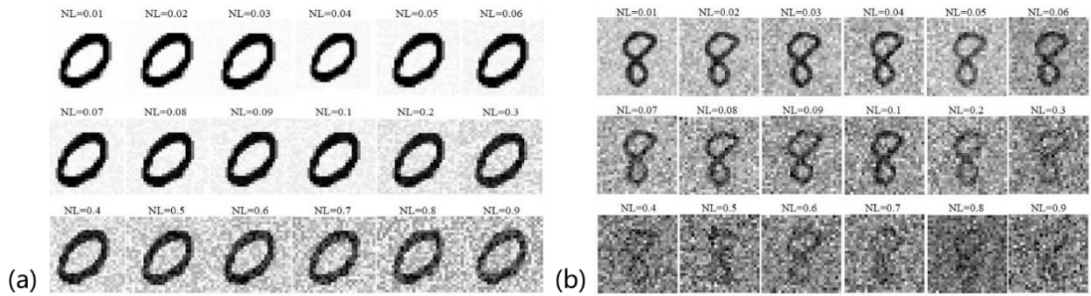

**Figure 8:** Datasets after adding noise. (a) (b) represents the addition of uniform and Gaussian noise, NL denotes the noise level.

### 4.2 Noise Comparison Results

In our noise experiments, we compared our network with SFNN and SQNN, as illustrated in Figure 9, unveiling a noteworthy superiority of our network. Our network consistently achieves higher accuracy than the other two networks across various levels of uniform and Gaussian noise. Notably, as the noise level increases, the advantages in accuracy become more pronounced. This phenomenon can be attributed to the unique parallel structure of our PPF-SQNN network, which effectively integrates the strengths of classical neural networks and quantum neural networks, showcasing exceptional performance in noisy environments.

Compared to the SFNN and SQNN, our parallel architecture that combines classical and quantum neural network components, resulting in enhanced capabilities and performance for image classifcation. Notably, the network excels in swiftly and accurately identifying and filtering noise signals, attributing its proficiency to both the efficient data processing inherent in classical neural networks and the integration of quantum properties within the quantum neural network. This parallel structure ensures superior adaptability to various levels and types of noise interference, leading to more stable and accurate classification outcomes. The network stands out in precisely differentiating between noise and authentic signals, maintaining a lower misjudgment rate. The classical neural network component plays a key role in feature extraction, while the quantum neural network component enhances robustness and feature extraction capacity during information processing [16]. The dual-component framework significantly elevates the overall classification performance of the network.

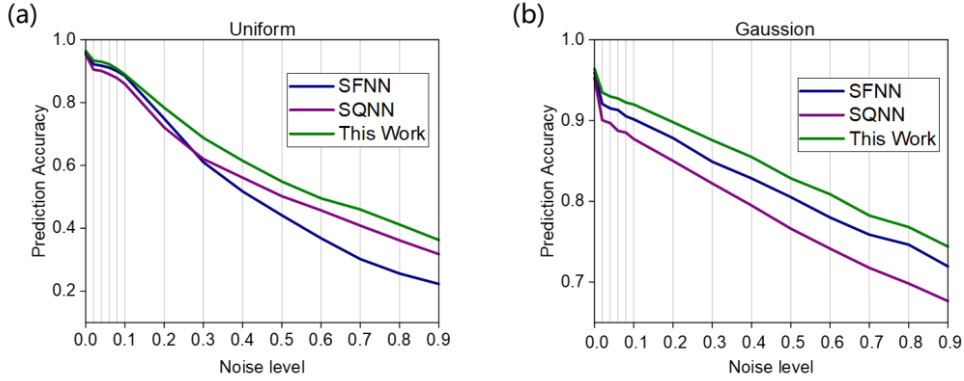

**Figure 9:** Comparison of prediction accuracy after adding noise.

5. Discussion

In our comparative analysis of experimental outcomes, we observe a notable proficiency of our network in managing extensive datasets. Primarily, both our training and testing accuracy surpass those of SFNN and SQNN significantly. This disparity underscores the heightened precision of our network in executing classification tasks. Furthermore, a detailed evaluation of model performance, utilizing confusion matrices, reinforces our network's enhanced reliability and superior capacity to distinguish between various sample categories.

To assess the robustness of our network, we conducted additional experiments involving the introduction of Uniform and Gaussian noise. The findings reveal substantial advantages of our network in noise-related experiments. In contrast to SFNN and SQNN, our network exhibits superior maintenance of classification accuracy and stability amid noise, affirming its heightened resilience to interference. This outcome suggests that the composite architecture, combining classical and quantum neural networks, holds considerable practical utility in mitigating noise challenges within real-world environments.

The reasons behind the advantages of our network warrants in-depth exploration and research. Firstly, classical neural networks and quantum neural networks have different characteristics and strengths, complementing each other in information processing and pattern recognition. The parallel architecture adeptly leverages these strengths, enhancing classification accuracy. Secondly, our network utilizes a series of effective optimization algorithms and parameter settings during the training process, resulting in better generalization and expressive capability of the network. Finally, our network exhibits better robustness in noisy environments, presumably owing to specific properties inherent in quantum neural networks, such as quantum entanglement [35] and quantum superposition, enabling the network to better handle noise interference.

Additionally, our research aims to delve into the scalability, efficiency, and adaptability of the network, particularly in handling more intricate tasks. The proposed approach serves as a versatile methodology for amalgamating features from classical neural networks and quantum neural networks, as depicted in Figure 10. This involves the computation of feature vectors derived by both networks for image feature extraction, utilizing Equation (12) to allocate network weights and generate the fusion feature vector. Importantly, this methodology extends beyond the scope of spiking neural networks. Given the interconnected nature of classical neural networks, a variety of traditional structures, including convolutional neural networks (CNNs) [11] for image feature extraction and recurrent neural networks (RNNs) [36] for sequence data processing, can be seamlessly integrated. Such flexibility empowers us to select the most suitable classical neural network structure for efficient integration with quantum neural networks across diverse fields and tasks. Moreover, our proposed architecture is not confined to the parallel fusion of two networks; it can be extended to incorporate multiple networks, thereby creating a more potent and diversified fused model.

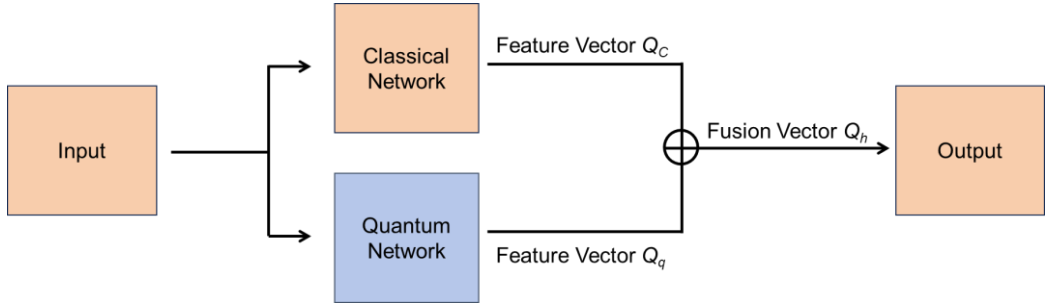

**Figure 10:** A general architecture for the parallel fusion of classical and quantum neural networks.

6. Conclusion

We introduce a pioneering PPF-SQNN architecture for image classification, leveraging a parallel proportional fusion strategy that combines classical Spiking Neural Networks and Variational Quantum Circuits. The research provides notable contributions to the field, including the introduction of an innovative network weight optimization algorithm, parametrization of the network, and a thorough evaluation of its performance in the presence of noise. The key findings suggest that PPF-SQNN exhibits superior accuracy, robustness, and noise immunity compared to classical SFNN and serial hybrid SQNN architectures. The experimental results highlight the significance of the parallel fusion architecture, effectively harnessing the complementary strengths of classical and quantum neural networks. The proposed network not only surpasses existing models in terms of accuracy but also demonstrates exceptional stability and adaptability in noisy environments. The optimization techniques employed, coupled with the network's unique structure, contribute to its heightened generalization and expressive capabilities.

Moreover, our work emphasizes the versatility and scalability of the proposed approach, allowing for seamless integration with various classical neural network structures beyond SNN, such as CNNs and RNNs. This adaptability positions PPF-SQNN as a promising candidate for diverse applications beyond image classification. Looking ahead, future research directions include exploring the scalability and adaptability of the network for more complex tasks, extending its applicability to real-world problems in diverse domains, and delving deeper into the specific quantum properties that contribute to its robustness. Ongoing investigations into optimization techniques and algorithmic enhancements aim to further enhance the efficiency and performance of the hybrid network. The comprehensive and innovative nature of this work underscores its potential to advance the field of quantum-spiking neural networks and its practical utility in addressing challenges in real-world applications. Our network holds tremendous potential for addressing complex classification problems, bringing new opportunities and challenges to the development of related fields.

Acknowledgments

This work was partly supported by the National Natural Science Foundation of China (Grant Nos. 61874001, 62004001, 62201005, 62202003, 62004001, 62304001), the Anhui Provincial Natural Science Foundation under Grant No. 2308085QF213, No. 2108085MF228, and the Natural Science Research Project of Anhui Educational Committee under Grant No. 2022AH050106, 2023AH050072.

References

[1] Yanming, Guo, Yu, Liu, Ard, Oerlemans, Songyang, Lao, Song, Wu, Deep learning for visual understanding: A review, Neurocomputing 187(Apr.26) (2016) 27-48.

[2] Y. Chen, C. Wang, H. Guo, X. Gao, J. Wu, Accelerating spiking neural networks using quantum algorithm with high success probability and high calculation accuracy, Neurocomputing 493 (2022) 435-444.

[3] D. Bokhan, A.S. Mastiukova, A.S. Boev, D.N. Trubnikov, A.K. Fedorov, Multiclass classification using quantum convolutional neural networks with hybrid quantum-classical learning (vol 10, 1069985, 2022), Frontiers in Physics 10 (2023).

[4] S. Alsubai, A. Alqahtani, A. Binbusayyis, M. Sha, A. Gumaei, S.H. Wang, A Quantum Computing-Based Accelerated Model for Image Classification Using a Parallel Pipeline Encoded Inception Module, Mathematics 11(11) (2023).

[5] S.S. Shi, Z.M. Wang, R.M. Shang, Y.A. Li, J.X. Li, G.Q. Zhong, Y.J. Gu, Hybrid quantum-classical convolutional neural network for phytoplankton classification, Frontiers in Marine Science 10 (2023).

[6] A.J. Wang, J.L. Hu, S.Y. Zhang, L.S. Li, Shallow hybrid quantum-classical convolutional neural network model for image classification, Quantum Information Processing 23(1) (2024).

[7] L. Domingo, M. Djukic, C. Johnson, F. Borondo, Binding affinity predictions with hybrid quantum-classical convolutional neural networks, Scientific Reports 13(1) (2023) 17951.

[8] M. Schuld, I. Sinayskiy, F. Petruccione, Simulating a perceptron on a quantum computer, Physics Letters A 379(7) (2015) 660-663.

[9] A. Chalumuri, R. Kune, S. Kannan, B.S. Manoj, Quantum-enhanced deep neural network architecture for image scene classification, Quantum Information Processing 20(11) (2021).

[10] J. Preskill, Quantum Computing in the NISQ era and beyond, Quantum 2 (2018).

[11] Y. LeCun, B. Boser, J.S. Denker, D. Henderson, R.E. Howard, W. Hubbard, L.D. Jackel, Backpropagation Applied to Handwritten Zip Code Recognition, Neural Computation 1(4) (1989) 541-551.

[12] D.H. Ackley, G.E. Hinton, T.J. Sejnowski, A learning algorithm for boltzmann machines, Cognitive Science 9(1) (1985) 147-169.

[13] I. Goodfellow, J. Pouget-Abadie, M. Mirza, B. Xu, D. Warde-Farley, S. Ozair, A. Courville, Y. Bengio, Generative Adversarial Nets, Neural Information Processing Systems, 2014.

[14] S.J. Pan, Q.A. Yang, A Survey on Transfer Learning, Ieee Transactions on Knowledge and Data Engineering 22(10) (2010) 1345-1359.

[15] Y. Sun, Y. Zeng, T. Zhang, Quantum superposition inspired spiking neural network, iScience 24(8) (2021) 102880.

[16] D. Konar, A.D. Sarma, S. Bhandary, S. Bhattacharyya, A. Cangi, V. Aggarwal, A shallow hybrid classical–quantum spiking feedforward neural network for noise-robust image classification, Applied Soft Computing 136 (2023) 110099.

[17] M. Kordzanganeh, D. Kosichkina, A. Melnikov, Parallel Hybrid Networks: An Interplay between Quantum and Classical Neural Networks, Intelligent Computing 2 (2023) 0028.

[18] C. Szegedy, L. Wei, J. Yangqing, P. Sermanet, S. Reed, D. Anguelov, D. Erhan, V. Vanhoucke, A. Rabinovich, Going deeper with convolutions, 2015 IEEE Conference on Computer Vision and Pattern Recognition (CVPR), 2015, pp. 1-9.

[19] A. Zhang, H. Zhou, X. Li, W. Zhu, Fast and robust learning in Spiking Feed-forward Neural Networks based on Intrinsic Plasticity mechanism, Neurocomputing 365 (2019) 102-112.

[20] Y. Cao, Y. Chen, D. Khosla, Spiking Deep Convolutional Neural Networks for Energy-Efficient Object Recognition, International Journal of Computer Vision 113(1) (2015) 54-66.

[21] M. Prezioso, M.R. Mahmoodi, F.M. Bayat, H. Nili, H. Kim, A. Vincent, D.B. Strukov, Spike-timing-dependent plasticity learning of coincidence detection with passively integrated memristive circuits, Nature Communications 9(1) (2018) 5311.

[22] N. Srivastava, G. Hinton, A. Krizhevsky, I. Sutskever, R. Salakhutdinov, Dropout: a simple way to prevent neural networks from overfitting, J. Mach. Learn. Res. 15(1) (2014) 1929–1958.

[23] V. Nair, G.E. Hinton, Rectified Linear Units Improve Restricted Boltzmann Machines, International Conference on Machine Learning, 2010.

[24] A. Peruzzo, J. McClean, P. Shadbolt, M.H. Yung, X.Q. Zhou, P.J. Love, A. Aspuru-Guzik, J.L. O'Brien, A variational eigenvalue solver on a photonic quantum processor, Nat Commun 5 (2014) 4213.

[25] J. Qi, C.-H.H. Yang, P.-Y. Chen, M.-H. Hsieh, Theoretical error performance analysis for variational quantum circuit based functional regression, npj Quantum Information 9 (2022) 1-10.

[26] M. Benedetti, E. Lloyd, S. Sack, M. Fiorentini, Parameterized quantum circuits as machine learning models, Quantum Science and Technology 4(4) (2019) 043001.

[27] P.Q. Le, A.M. Iliyasu, F. Dong, K. Hirota, A flexible representation of quantum images for polynomial preparation, image compression and processing operations, Quantum Inf, Quantum Information Processing 10(1) (2011) 63-84.

[28] D. Konar, S. Bhattacharyya, B.K. Panigrahi, E.C. Behrman, Qutrit-Inspired Fully Self-Supervised Shallow Quantum Learning Network for Brain Tumor Segmentation, IEEE Trans Neural Netw Learn Syst 33(11) (2022) 6331-6345.

[29] D. Konar, S. Bhattacharyya, T.K. Gandhi, B.K. Panigrahi, A Quantum-Inspired Self-Supervised Network model for automatic segmentation of brain MR images, Applied Soft Computing 93 (2020) 106348.

[30] Y. Lecun, L. Bottou, Gradient-based learning applied to document recognition, Proceedings of the IEEE 86(11) (1998) 2278-2324.

[31] S.R. Kheradpisheh, M. Ganjtabesh, S.J. Thorpe, T. Masquelier, STDP-based spiking deep convolutional neural networks for object recognition, Neural Networks 99 (2018) 56-67.

[32] Peter, U., Diehl, Matthew, Cook, Unsupervised learning of digit recognition using spike-timing-dependent plasticity, Frontiers in Computational Neuroscience 9 (2015) 99.

[33] B. Zhao, R. Ding, S. Chen, B. Linares-Barranco, H. Tang, Feedforward Categorization on AER Motion Events Using Cortex-Like Features in a Spiking Neural Network, IEEE Trans Neural Netw Learn Syst 26(9) (2015) 1963-78.

[34] L. Wan, M. Zeiler, S. Zhang, Y. LeCun, R. Fergus, Regularization of neural networks using dropconnect, Proceedings of the 30th International Conference on International Conference on Machine Learning - Volume 28, JMLR.org, Atlanta, GA, USA, 2013, pp. III–1058–III–1066.

[35] F.J. Duarte, Fundamentals of Quantum Entanglement, IOP Publishing, 2019.

[36] S. Hochreiter, J. Schmidhuber, Long Short-Term Memory, Neural Computation 9(8) (1997) 1735-1780.